\newif\ifprd
\prdfalse
\ifprd
\documentclass[%
aps,%
prd,%
showpacs,%
floatfix,
amsmath,
amssymb,
eqsecnum,%
twocolumn,%
nofootinbib%
]{revtex4}
\else
\documentclass[12pt]{article}
\fi

\newif\ifglobal
\globaltrue

\usepackage{ifthen}
\usepackage{calc}
\ifprd{}
\else
	\usepackage{amsmath}
	\usepackage{amssymb}
	\usepackage{url}
\fi
\usepackage{graphicx}	
\usepackage{mathrsfs}	
\usepackage{bbm}		
\usepackage{natbib}	
\usepackage[usenames]{color}	

\newcommand{\ourtitle}{Minimal Seesaw as an Ultraviolet Insensitive Cure for the Problems of Anomaly Mediation}
\newcommand{\paperauth}{R. N. Mohapatra, N. Setzer, and S. Spinner}
\newcommand{\ouraffiliation}{Maryland Center for Fundamental Physics and Department of Physics, University of Maryland, College Park, MD 20742, USA}
\newcommand{\ourpreprint}{\vbox{ \hbox{UMD-PP-XX-XXX} }}
\newcommand{\ourpacs}{14.60.Pq, 98.80.Cq}
\newcommand{\ourdate}{}

\ifprd{}
\else
\title{\ourtitle}
\author{\paperauth}	
\fi

\newcommand{\pderiv}[2]{ \frac{\partial {#1}}{\partial {#2}} }

\newcommand{\deriv}[2]{   \frac{ d {#1}}{ d {#2} }  }

\newcommand{\intOp}[2][]{\int \! d^{#1}#2 \;}

\newcommand{\abs}[1]{ \left| \, {#1} \right| }

\DeclareMathOperator{\sgn}{sgn}

\newcommand{\half}{\frac{1}{2}}
\newcommand{\third}{\frac{1}{3}}

\DeclareMathOperator{\Tr}{Tr}

\newcommand{\inp}[2][0cm]{ \mathopen{}\left( #2 \parbox[h][#1]{0cm}{} \right) }
\newcommand{\inb}[2][0cm]{ \mathopen{}\left[ #2 \parbox[h][#1]{0cm}{} \right] }

\newcommand{\inap}[2][0cm]{ \mathopen{}\left< {#2} \parbox[h][#1]{0cm}{} \right> }
\newcommand{\nop}[1]{\mathopen{}\left.{#1}\right.}
\newcommand{\pfrac}[2]{ \mathopen{}\left( \frac{#1}{#2} \right) }

\newcommand{\eat}[1]{}
\newcommand{\eq}[1]{Eq.~\eqref{Eq:#1}}
\newcommand{\eqn}[1]{\eqref{Eq:#1}}
\newcommand{\fig}[1]{Figure~\ref{Fig:#1}}
\newcommand{\Sec}[1]{Section~\ref{Sec:#1}}
\newcommand{\tbl}[1]{Table~\ref{Table:#1}}
\newcommand{\app}[1]{Appendix~\ref{App:#1}}
\newcounter{itemnum}
\newenvironment{olist}
{%
\begin{list}%
	{(\roman{itemnum})}%
	{
	\usecounter{itemnum}}%
}%
{%
\end{list}
}%

\let\gobbleEnd=\relax
\newcommand{\drawmasslines}[2]{\gobblemasslist#2,\gobbleHalt\gobbleEnd}
\newcommand{\gobbleHalt}{\@\@\@\@}

\def\gobblemasslist(#1,#2,[#3,#4]),#5\gobbleEnd{%
\ifthenelse{ \equal{#5}{\gobbleHalt} }%
	{%
	\put(\xCoord,#2){\line(1,0){\masslinelen} \raisebox{#4}{\hspace{#3}$#1$}}%
	}%
	{%
	\put(\xCoord,#2){\line(1,0){\masslinelen} \raisebox{#4}{\hspace{#3}$#1$}}%
	\gobblemasslist#5\gobbleEnd%
	}%
}%

\newlength{\oldunitlength}
\newcommand{\masslinelen}{50}
\newcommand{\xCoordColOne}{10}
\newcommand{\xCoordColTwo}{120}
\newcommand{\xCoordColThree}{230}
\newcommand{\xCoordColFour}{340}
\newcommand{\xCoordColFive}{450}
\newcommand{\xCoord}{}
\newcommand{\defaultX}{1mm}
\newcommand{\defaultY}{-0.6ex}
\newcommand{\placetoleft}{-1.4cm}
\newcommand{\placetoleftnudgeright}{-1.33cm}
\newcommand{\slightlyraised}{-0.4ex}
\newcommand{\slightlylowered}{-0.8ex}

\newcommand{\E}[1]{\times 10^{#1}}	
\newcommand{\vev}[1]{\inap{#1}}		

\newcommand{\MP}{M_{\text{Pl}}} 		
\newcommand{\MSUSY}{M_\text{SUSY}}	
\newcommand{\Mmess}{M_\text{mess}}	

\newcommand{\G}[1]{\ensuremath{G_{#1}}}	
\newcommand{\SUSYLR}{SUSYLR}
\newcommand{\CISLR}{CISLR}

\newcommand{\WSUSYLR}{W_{\text{\SUSYLR}}}

\newcommand{\ourmodel}{AMSBLR}

\newcommand{\DeltaBar}{\bar{\Delta}}

\newcommand{\DeltaC}{\Delta^c}
\newcommand{\DeltaBarC}{\bar{\Delta}^c}

\newcommand{\scalar}[1]{%
\ifthenelse{ \equal{#1}{\DeltaC} }%
	{%
	\underline{\Delta}^c%
	}%
{ \ifthenelse{ \equal{#1}{\DeltaBarC} }%
	{%
	\underline{\bar{\Delta}}^c%
	}%
{ \ifthenelse{ \equal{#1}{\Phi} }%
	{%
	\underline{\Phi}%
	}%
	{%
	\underline{#1}%
	}}}%
}%

\newcommand{\DBarCpp}{\bar{\Delta}^{c ++}}
\newcommand{\DBarCp}{\bar{\Delta}^{c +}}
\newcommand{\DBarCz}{\bar{\Delta}^{c \,0}}
\newcommand{\DCmm}{\Delta^{c --}}
\newcommand{\DCm}{\Delta^{c -}}
\newcommand{\DCz}{\Delta^{c \,0}}
\newcommand{\phiUp}{\Phi_u^+}
\newcommand{\phiUz}{\Phi_u^0}
\newcommand{\phiDz}{\Phi_d^0}
\newcommand{\phiDm}{\Phi_d^-}

\newcommand{\Slight}{N}
\newcommand{\Sheavy}{S}

\newcommand{\kappaLight}{\kappa_\Slight}
\newcommand{\kappaHeavy}{\kappa_\Sheavy}

\newcommand{\lambdaLight}{\lambda_\Slight}
\newcommand{\lambdaHeavy}{\lambda_\Sheavy}

\newcommand{\I}{\mathbbm{i}}
\newcommand{\EE}{\mathbbm{e}}

\newcommand{\bibpath}{.}
\begin{document}

\ifprd
\preprint{\ourpreprint}
\title{\ourtitle}
\author{\paperauth}	
\affiliation{\ouraffiliation}
\date{\ourdate}
\pacs{\ourpacs}
\else
\begin{center}
\huge \ourtitle

\vspace{0.5cm}
\large
\paperauth

\normalsize
\textit{\ouraffiliation} 

\vspace{0.5cm}

\large
\today 

\end{center}
\vspace{1cm}
\fi

\begin{abstract}
\makeatletter%
\@ifundefined{ifglobal}%
{%
\newif\ifglobal%
\globalfalse%
\newif\ifprd%
\prdfalse%
}%
\makeatother%
\ifglobal{}%
\else%
\documentclass[12pt]{article}

\usepackage{ifthen}
\usepackage{calc}
\ifprd{}
\else
	\usepackage{amsmath}
	\usepackage{amssymb}
	\usepackage{url}
\fi
\usepackage{graphicx}	
\usepackage{mathrsfs}	
\usepackage{bbm}		
\usepackage{natbib}	
\usepackage[usenames]{color}	

\newcommand{\ourtitle}{Minimal Seesaw as an Ultraviolet Insensitive Cure for the Problems of Anomaly Mediation}
\newcommand{\paperauth}{R. N. Mohapatra, N. Setzer, and S. Spinner}
\newcommand{\ouraffiliation}{Maryland Center for Fundamental Physics and Department of Physics, University of Maryland, College Park, MD 20742, USA}
\newcommand{\ourpreprint}{\vbox{ \hbox{UMD-PP-XX-XXX} }}
\newcommand{\ourpacs}{14.60.Pq, 98.80.Cq}
\newcommand{\ourdate}{}

\ifprd{}
\else
\title{\ourtitle}
\author{\paperauth}	
\fi


\newcommand{\eq}[1]{Eq.~\eqref{Eq:#1}}
\newcommand{\eqn}[1]{\eqref{Eq:#1}}
\newcommand{\fig}[1]{Figure~\ref{Fig:#1}}
\newcommand{\Sec}[1]{Section~\ref{Sec:#1}}
\newcommand{\tbl}[1]{Table~\ref{Table:#1}}
\newcommand{\app}[1]{Appendix~\ref{App:#1}}
\newcounter{itemnum}
\newenvironment{olist}
{%
\begin{list}%
	{(\roman{itemnum})}%
	{
	\usecounter{itemnum}}%
}%
{%
\end{list}
}%

\let\gobbleEnd=\relax
\newcommand{\drawmasslines}[2]{\gobblemasslist#2,\gobbleHalt\gobbleEnd}
\newcommand{\gobbleHalt}{\@\@\@\@}

\def\gobblemasslist(#1,#2,[#3,#4]),#5\gobbleEnd{%
\ifthenelse{ \equal{#5}{\gobbleHalt} }%
	{%
	\put(\xCoord,#2){\line(1,0){\masslinelen} \raisebox{#4}{\hspace{#3}$#1$}}%
	}%
	{%
	\put(\xCoord,#2){\line(1,0){\masslinelen} \raisebox{#4}{\hspace{#3}$#1$}}%
	\gobblemasslist#5\gobbleEnd%
	}%
}%

\newlength{\oldunitlength}
\setlength{\oldunitlength}{\the\unitlength}
\newcommand{\masslinelen}{50}
\newcommand{\xCoordColOne}{10}
\newcommand{\xCoordColTwo}{120}
\newcommand{\xCoordColThree}{230}
\newcommand{\xCoordColFour}{340}
\newcommand{\xCoordColFive}{450}
\newcommand{\xCoord}{}
\newcommand{\defaultX}{1mm}
\newcommand{\defaultY}{-0.6ex}
\newcommand{\placetoleft}{-1.4cm}
\newcommand{\placetoleftnudgeright}{-1.33cm}
\newcommand{\slightlyraised}{-0.4ex}
\newcommand{\slightlylowered}{-0.8ex}


\newcommand{\E}[1]{\times 10^{#1}}	
\newcommand{\vev}[1]{\inap{#1}}		

\newcommand{\MP}{M_{\text{Pl}}} 		
\newcommand{\MSUSY}{M_\text{SUSY}}	
\newcommand{\Mmess}{M_\text{mess}}	

\newcommand{\G}[1]{\ensuremath{G_{#1}}}	
\newcommand{\SUSYLR}{SUSYLR}
\newcommand{\CISLR}{CISLR}

\newcommand{\WSUSYLR}{W_{\text{\SUSYLR}}}

\newcommand{\ourmodel}{AMSBLR}

\newcommand{\DeltaBar}{\bar{\Delta}}

\newcommand{\DeltaC}{\Delta^c}
\newcommand{\DeltaBarC}{\bar{\Delta}^c}

\newcommand{\scalar}[1]{%
\ifthenelse{ \equal{#1}{\DeltaC} }%
	{%
	\underline{\Delta}^c%
	}%
{ \ifthenelse{ \equal{#1}{\DeltaBarC} }%
	{%
	\underline{\bar{\Delta}}^c%
	}%
{ \ifthenelse{ \equal{#1}{\Phi} }%
	{%
	\underline{\Phi}%
	}%
	{%
	\underline{#1}%
	}}}%
}%

\newcommand{\DBarCpp}{\bar{\Delta}^{c ++}}
\newcommand{\DBarCp}{\bar{\Delta}^{c +}}
\newcommand{\DBarCz}{\bar{\Delta}^{c \,0}}
\newcommand{\DCmm}{\Delta^{c --}}
\newcommand{\DCm}{\Delta^{c -}}
\newcommand{\DCz}{\Delta^{c \,0}}
\newcommand{\phiUp}{\Phi_u^+}
\newcommand{\phiUz}{\Phi_u^0}
\newcommand{\phiDz}{\Phi_d^0}
\newcommand{\phiDm}{\Phi_d^-}

\newcommand{\Slight}{N}
\newcommand{\Sheavy}{S}

\newcommand{\kappaLight}{\kappa_\Slight}
\newcommand{\kappaHeavy}{\kappa_\Sheavy}

\newcommand{\lambdaLight}{\lambda_\Slight}
\newcommand{\lambdaHeavy}{\lambda_\Sheavy}

\newcommand{\I}{\mathbbm{i}}
\newcommand{\EE}{\mathbbm{e}}

\newcommand{\bibpath}{.}
\bibliographystyle{h-physrev}
\citestyle{plain}

\begin{document}

\maketitle

\section{Abstract}

\fi%

We show that an intermediate scale supersymmetric left-right seesaw scenario with automatic $R$-parity conservation can  cure the problem of tachyonic slepton masses that arises when supersymmetry is broken by anomaly mediation, while preserving ultraviolet insensitivity.  The reason for this is the existence of light $B-L=2$ higgses with yukawa couplings to the charged leptons.  We find these theories to have distinct predictions compared to the usual mSUGRA and gauge mediated models as well as the minimal AMSB models.  Such predictions include a condensed gaugino mass spectrum and possibly a correspondingly condensed sfermion spectrum.

\ifglobal{}%
\else%
%
\bibliography{\bibpath/PR}

\end{document}
\fi%
\end{abstract}

\ifprd
\maketitle
\fi

\section{Introduction}
\label{Sec:Intro}

\makeatletter%
\@ifundefined{ifglobal}%
{%
\newif\ifglobal%
\globalfalse%
\newif\ifprd%
\prdfalse%
}%
\makeatother%
\ifglobal{}%
\else%
\documentclass[12pt]{article}

\usepackage{ifthen}
\usepackage{calc}
\ifprd{}
\else
	\usepackage{amsmath}
	\usepackage{amssymb}
	\usepackage{url}
\fi
\usepackage{graphicx}	
\usepackage{mathrsfs}	
\usepackage{bbm}		
\usepackage{natbib}	
\usepackage[usenames]{color}	

\newcommand{\ourtitle}{Minimal Seesaw as an Ultraviolet Insensitive Cure for the Problems of Anomaly Mediation}
\newcommand{\paperauth}{R. N. Mohapatra, N. Setzer, and S. Spinner}
\newcommand{\ouraffiliation}{Maryland Center for Fundamental Physics and Department of Physics, University of Maryland, College Park, MD 20742, USA}
\newcommand{\ourpreprint}{\vbox{ \hbox{UMD-PP-XX-XXX} }}
\newcommand{\ourpacs}{14.60.Pq, 98.80.Cq}
\newcommand{\ourdate}{}

\ifprd{}
\else
\title{\ourtitle}
\author{\paperauth}	
\fi


\newcommand{\eq}[1]{Eq.~\eqref{Eq:#1}}
\newcommand{\eqn}[1]{\eqref{Eq:#1}}
\newcommand{\fig}[1]{Figure~\ref{Fig:#1}}
\newcommand{\Sec}[1]{Section~\ref{Sec:#1}}
\newcommand{\tbl}[1]{Table~\ref{Table:#1}}
\newcommand{\app}[1]{Appendix~\ref{App:#1}}
\newcounter{itemnum}
\newenvironment{olist}
{%
\begin{list}%
	{(\roman{itemnum})}%
	{
	\usecounter{itemnum}}%
}%
{%
\end{list}
}%

\let\gobbleEnd=\relax
\newcommand{\drawmasslines}[2]{\gobblemasslist#2,\gobbleHalt\gobbleEnd}
\newcommand{\gobbleHalt}{\@\@\@\@}

\def\gobblemasslist(#1,#2,[#3,#4]),#5\gobbleEnd{%
\ifthenelse{ \equal{#5}{\gobbleHalt} }%
	{%
	\put(\xCoord,#2){\line(1,0){\masslinelen} \raisebox{#4}{\hspace{#3}$#1$}}%
	}%
	{%
	\put(\xCoord,#2){\line(1,0){\masslinelen} \raisebox{#4}{\hspace{#3}$#1$}}%
	\gobblemasslist#5\gobbleEnd%
	}%
}%

\newlength{\oldunitlength}
\setlength{\oldunitlength}{\the\unitlength}
\newcommand{\masslinelen}{50}
\newcommand{\xCoordColOne}{10}
\newcommand{\xCoordColTwo}{120}
\newcommand{\xCoordColThree}{230}
\newcommand{\xCoordColFour}{340}
\newcommand{\xCoordColFive}{450}
\newcommand{\xCoord}{}
\newcommand{\defaultX}{1mm}
\newcommand{\defaultY}{-0.6ex}
\newcommand{\placetoleft}{-1.4cm}
\newcommand{\placetoleftnudgeright}{-1.33cm}
\newcommand{\slightlyraised}{-0.4ex}
\newcommand{\slightlylowered}{-0.8ex}


\newcommand{\E}[1]{\times 10^{#1}}	
\newcommand{\vev}[1]{\inap{#1}}		

\newcommand{\MP}{M_{\text{Pl}}} 		
\newcommand{\MSUSY}{M_\text{SUSY}}	
\newcommand{\Mmess}{M_\text{mess}}	

\newcommand{\G}[1]{\ensuremath{G_{#1}}}	
\newcommand{\SUSYLR}{SUSYLR}
\newcommand{\CISLR}{CISLR}

\newcommand{\WSUSYLR}{W_{\text{\SUSYLR}}}

\newcommand{\ourmodel}{AMSBLR}

\newcommand{\DeltaBar}{\bar{\Delta}}

\newcommand{\DeltaC}{\Delta^c}
\newcommand{\DeltaBarC}{\bar{\Delta}^c}

\newcommand{\scalar}[1]{%
\ifthenelse{ \equal{#1}{\DeltaC} }%
	{%
	\underline{\Delta}^c%
	}%
{ \ifthenelse{ \equal{#1}{\DeltaBarC} }%
	{%
	\underline{\bar{\Delta}}^c%
	}%
{ \ifthenelse{ \equal{#1}{\Phi} }%
	{%
	\underline{\Phi}%
	}%
	{%
	\underline{#1}%
	}}}%
}%

\newcommand{\DBarCpp}{\bar{\Delta}^{c ++}}
\newcommand{\DBarCp}{\bar{\Delta}^{c +}}
\newcommand{\DBarCz}{\bar{\Delta}^{c \,0}}
\newcommand{\DCmm}{\Delta^{c --}}
\newcommand{\DCm}{\Delta^{c -}}
\newcommand{\DCz}{\Delta^{c \,0}}
\newcommand{\phiUp}{\Phi_u^+}
\newcommand{\phiUz}{\Phi_u^0}
\newcommand{\phiDz}{\Phi_d^0}
\newcommand{\phiDm}{\Phi_d^-}

\newcommand{\Slight}{N}
\newcommand{\Sheavy}{S}

\newcommand{\kappaLight}{\kappa_\Slight}
\newcommand{\kappaHeavy}{\kappa_\Sheavy}

\newcommand{\lambdaLight}{\lambda_\Slight}
\newcommand{\lambdaHeavy}{\lambda_\Sheavy}

\newcommand{\I}{\mathbbm{i}}
\newcommand{\EE}{\mathbbm{e}}

\newcommand{\bibpath}{.}
\bibliographystyle{h-physrev}
\citestyle{plain}

\begin{document}

\maketitle

\section{Introduction}
\label{Sec:Intro}

\fi%

The driving motivations for physics beyond the standard model are: (i) stabilizing the Higgs mass against radiative corrections, thereby providing a firmer understanding of the origin of mass; (ii) understanding the neutrino masses and mixings, especially the extreme smallness of its mass compared to those of charged fermions; (iii) finding a particle physics candidate for the dark matter of the universe and (iv) solving the strong CP problem. Two prevalent ideas for resolving them are: supersymmetry (SUSY)---curing (i), and the seesaw mechanism\cite{Minkowski:1977sc, Gell-Mann:1980vs, Yanagida:1979as, Glashow:1979nm, Mohapatra:1979ia}---curing (ii)---making SUSY seesaw very enticing.  $R$-parity is assured as an automatic symmetry of the low energy lagrangian\cite{Mohapatra:1986su, Font:1989ai, Martin:1992mq} given $B-L$ is a gauged symmetry broken by $B-L = 2$ higgs fields.  Conservation of $R$-parity would guarantee a stable lightest SUSY particle (LSP),providing a good dark matter candidate\cite{Martin:1997ns} as well as preventing catastrophic proton decay (caused by $R$-parity breaking terms of the minimal SUSY standard model (MSSM)).  Finally, gauged $B-L$ models embedded into the SUSY left-right framework, provide a cure to the strong CP problem without the need for an axion\cite{Mohapatra:1995xd, Kuchimanchi:1995rp}. This leads us to focus on the minimal SUSY left-right model and look at further constraints when the method of SUSY breaking is considered.  

The nature and origin of SUSY breaking has been a focus of a great deal of attention. The detailed nature of new physics that breaks SUSY is unknown---although there are several interesting suggestions\cite{Martin:1997ns}. Here we focus on SUSY breaking is via anomaly mediation which is related to the radiative breaking of conformal SUSY\cite{Randall:1998uk, Giudice:1998xp}. Anomaly mediated SUSY breaking (AMSB) predicts all the soft SUSY breaking parameters in terms of one mass parameter (the mass of the gravitino) and the $\beta$ and $\gamma$ functions of the low energy theory.  As such, it is a highly predictive scenario which avoids the SUSY flavor problem (no new flavor physics is introduced) and solved the gravitino mass problem.

There is, however, a serious problem that afflicts any AMSB model whose low energy theory is the MSSM: the sleptons have negative mass squared thereby leading to a ground state that breaks electric charge. Finding a cure to this problem is a difficult task given the predictability of AMSB and the fact that it generally decouples higher scale physics.  This forces solutions to include new couplings in the low energy theory or deflecting the AMSB trajectories.  While proposed solutions along these lines \cite{Chacko:1999am, Gherghetta:1999sw, Katz:1999uw, Pomarol:1999ie, Arkani-Hamed:2000xj, Allanach:2000gu, Jack:2000cd, Carena:2000ad, Ibe:2004tg, Lebedev:2005ge} are very illuminating they lack an independent motive.  

In this paper we propose a new way to resolve this problem of AMSB using the minimal $R$-parity conserving SUSYLR seesaw model mentioned above. We present an instance of this class of bottom up seesaw models that has an intermediate seesaw scale (of order $10^{11}$ GeV or so) and show that the slepton mass square problem of AMSB is cured.  Furthermore, ultraviolet (UV) insensitivity is preserved; a featured that is shared with only a few of the proposed AMSB solutions.  The key to this is the existence of light doubly-charged higgses\cite{Chacko:1997cm} and light left-handed triplets and their yukawa couplings to the lepton superfields. The effects of these doubly-charged fields can be discovered in low energy experiments as they lead to characteristic mass predictions which are different from those of other SUSY breaking scenarios.  We will demostrate these differences between our model, minimal supergravity (mSUGRA), minimal gauge mediated SUSY breaking (mGMSB) and AMSB with a universal scalar mass addition $m_0^2$ (mAMSB).  Apart from experimental testability, a novel feature of our suggestion is that the cure is motivated from independent considerations.  These defining phenomenological conditions:
\begin{olist}
\item SUSY

\item  local $B-L$ symmetry as part of the gauge group $SU(2)_L\times SU(2)_R \times U(1)_{B-L}$ so that one can
implement the seesaw mechanism

\item $B-L$ symmetry breaking is such that it leaves $R$-parity unbroken and assuring that there is a naturally stable dark matter candidate

\item SUSY is broken radiatively by conformal anomalies, hence keeping the soft terms (renormalization group equation) RGE invariant down to the TeV scale (UV insensitivity).

\end{olist}

We will show in \Sec{SUSYLR} how these consideration produce slepton positive mass-squares, as well as introduce the model and gives its sparticle spectrum.  \Sec{AMSB} will give a brief overview of AMSB and introduce its notation---and terminology---a task to which we now turn.

\ifglobal{}%
\else%
%
\bibliography{\bibpath/Intro}

\end{document}
\fi%

\section{Summary of AMSB}
\label{Sec:AMSB}

\makeatletter%
\@ifundefined{ifglobal}%
{%
\newif\ifglobal%
\globalfalse%
\newif\ifprd%
\prdfalse%
}%
\makeatother%
\ifglobal{}%
\else%
\documentclass[12pt]{article}

\usepackage{ifthen}
\usepackage{calc}
\ifprd{}
\else
	\usepackage{amsmath}
	\usepackage{amssymb}
	\usepackage{url}
\fi
\usepackage{graphicx}	
\usepackage{mathrsfs}	
\usepackage{bbm}		
\usepackage{natbib}	
\usepackage[usenames]{color}	

\newcommand{\ourtitle}{Minimal Seesaw as an Ultraviolet Insensitive Cure for the Problems of Anomaly Mediation}
\newcommand{\paperauth}{R. N. Mohapatra, N. Setzer, and S. Spinner}
\newcommand{\ouraffiliation}{Maryland Center for Fundamental Physics and Department of Physics, University of Maryland, College Park, MD 20742, USA}
\newcommand{\ourpreprint}{\vbox{ \hbox{UMD-PP-XX-XXX} }}
\newcommand{\ourpacs}{14.60.Pq, 98.80.Cq}
\newcommand{\ourdate}{}

\ifprd{}
\else
\title{\ourtitle}
\author{\paperauth}	
\fi


\newcommand{\eq}[1]{Eq.~\eqref{Eq:#1}}
\newcommand{\eqn}[1]{\eqref{Eq:#1}}
\newcommand{\fig}[1]{Figure~\ref{Fig:#1}}
\newcommand{\Sec}[1]{Section~\ref{Sec:#1}}
\newcommand{\tbl}[1]{Table~\ref{Table:#1}}
\newcommand{\app}[1]{Appendix~\ref{App:#1}}
\newcounter{itemnum}
\newenvironment{olist}
{%
\begin{list}%
	{(\roman{itemnum})}%
	{
	\usecounter{itemnum}}%
}%
{%
\end{list}
}%

\let\gobbleEnd=\relax
\newcommand{\drawmasslines}[2]{\gobblemasslist#2,\gobbleHalt\gobbleEnd}
\newcommand{\gobbleHalt}{\@\@\@\@}

\def\gobblemasslist(#1,#2,[#3,#4]),#5\gobbleEnd{%
\ifthenelse{ \equal{#5}{\gobbleHalt} }%
	{%
	\put(\xCoord,#2){\line(1,0){\masslinelen} \raisebox{#4}{\hspace{#3}$#1$}}%
	}%
	{%
	\put(\xCoord,#2){\line(1,0){\masslinelen} \raisebox{#4}{\hspace{#3}$#1$}}%
	\gobblemasslist#5\gobbleEnd%
	}%
}%

\newlength{\oldunitlength}
\setlength{\oldunitlength}{\the\unitlength}
\newcommand{\masslinelen}{50}
\newcommand{\xCoordColOne}{10}
\newcommand{\xCoordColTwo}{120}
\newcommand{\xCoordColThree}{230}
\newcommand{\xCoordColFour}{340}
\newcommand{\xCoordColFive}{450}
\newcommand{\xCoord}{}
\newcommand{\defaultX}{1mm}
\newcommand{\defaultY}{-0.6ex}
\newcommand{\placetoleft}{-1.4cm}
\newcommand{\placetoleftnudgeright}{-1.33cm}
\newcommand{\slightlyraised}{-0.4ex}
\newcommand{\slightlylowered}{-0.8ex}


\newcommand{\E}[1]{\times 10^{#1}}	
\newcommand{\vev}[1]{\inap{#1}}		

\newcommand{\MP}{M_{\text{Pl}}} 		
\newcommand{\MSUSY}{M_\text{SUSY}}	
\newcommand{\Mmess}{M_\text{mess}}	

\newcommand{\G}[1]{\ensuremath{G_{#1}}}	
\newcommand{\SUSYLR}{SUSYLR}
\newcommand{\CISLR}{CISLR}

\newcommand{\WSUSYLR}{W_{\text{\SUSYLR}}}

\newcommand{\ourmodel}{AMSBLR}

\newcommand{\DeltaBar}{\bar{\Delta}}

\newcommand{\DeltaC}{\Delta^c}
\newcommand{\DeltaBarC}{\bar{\Delta}^c}

\newcommand{\scalar}[1]{%
\ifthenelse{ \equal{#1}{\DeltaC} }%
	{%
	\underline{\Delta}^c%
	}%
{ \ifthenelse{ \equal{#1}{\DeltaBarC} }%
	{%
	\underline{\bar{\Delta}}^c%
	}%
{ \ifthenelse{ \equal{#1}{\Phi} }%
	{%
	\underline{\Phi}%
	}%
	{%
	\underline{#1}%
	}}}%
}%

\newcommand{\DBarCpp}{\bar{\Delta}^{c ++}}
\newcommand{\DBarCp}{\bar{\Delta}^{c +}}
\newcommand{\DBarCz}{\bar{\Delta}^{c \,0}}
\newcommand{\DCmm}{\Delta^{c --}}
\newcommand{\DCm}{\Delta^{c -}}
\newcommand{\DCz}{\Delta^{c \,0}}
\newcommand{\phiUp}{\Phi_u^+}
\newcommand{\phiUz}{\Phi_u^0}
\newcommand{\phiDz}{\Phi_d^0}
\newcommand{\phiDm}{\Phi_d^-}

\newcommand{\Slight}{N}
\newcommand{\Sheavy}{S}

\newcommand{\kappaLight}{\kappa_\Slight}
\newcommand{\kappaHeavy}{\kappa_\Sheavy}

\newcommand{\lambdaLight}{\lambda_\Slight}
\newcommand{\lambdaHeavy}{\lambda_\Sheavy}

\newcommand{\I}{\mathbbm{i}}
\newcommand{\EE}{\mathbbm{e}}

\newcommand{\bibpath}{.}
\bibliographystyle{h-physrev}
\citestyle{plain}

\begin{document}

\maketitle

\section{AMSB}
\label{Sec:AMSB}

\fi%

AMSB has many attractive features: a large number of predictions, few parameters, an insensitivity to the UV and a mathematical framework that elegantly describes its affects.  The latter property allows one to express the SUSY breaking effects by analytically continuing parameters into superspace.  AMSB then gives a method or set of rules on how to ``promote" these parameters to superfields.  To establish these rules, as well as get the basic concepts of AMSB we start with a generic SUSY theory given by the lagrangian: 
\begin{equation}
\label{Eq:AMSB.Gen.SUSY.L}
\mathcal{ L}
	= \half \intOp[4]{\theta} \mathcal{K} \inp{D_\alpha, Q, W_\alpha} 
	  + \intOp[2]{\theta} \mathcal{W} \inp{Q, W_\alpha}
	  + \text{h.c.}
\end{equation}
where $Q$ collectively represents the matter content and $W_\alpha$ is the gauge content---the dependence of $\mathcal{K}$ on $\bar{D}_{\dot{\alpha}}$, $Q^\dagger$, etc.~has been suppressed.

AMSB then requires that $\mathcal{K}$ and $\mathcal{W}$ superconformal.  To do this it is necessary to introduce the superconformal compensator $\phi$ which is an unphysical (in that its scalar and fermionic components may be gauged away) chiral multiplet with a weyl weight $d_{W}(\phi) = +1$ and an $R$ charge of $+2/3$.  The superconformal invariance then dictates the $\phi$ couplings so that the resulting theory is invariant under weyl scale transformations and $U(1)_R$.

\begin{table}
\begin{center}
\begin{tabular}{|c|cc|}
\hline
\hline
		& $d_W$		& $R$		\\
\hline
$\theta$		& $-\half$	& $+1$		\\	
$\bar{\theta}$ 	& $-\half$	& $-1$		\\
$d \theta$ 	& $+\half$	& $-1$		\\
$d\bar{\theta}$	& $+\half$	& $+1$		\\
\hline
\hline
\end{tabular}%
\end{center}
\caption{Weyl weight and $R$ charges of superspace coordinates}
\label{Table:AMSB.theta.d.R.vals}%
\end{table}

\begin{table}
\begin{center}
\begin{tabular}{|c|cc|}
\hline
\hline
		& $d_W$		& $R$		\\
\hline
$\cal K$		& $+2$		& $0$		\\	
$\cal W$ 	& $+3$		& $+2$		\\
\hline
\hline
\end{tabular}%
\end{center}
\caption{Derived weyl weight and $R$ charge assignments for the K\"{a}hler and Super Potentials}
\label{Table:AMSB.K.W.d.R.vals}
\end{table}

To see the required form for the $\phi$ coupling, we first note that the superspace coordinate charge assignments (See \tbl{AMSB.theta.d.R.vals}) force the K\"ahler potential and Superpotential to have the charges shown in \tbl{AMSB.K.W.d.R.vals}.  If we take $d_W(\tilde{Q}) = d_W(\tilde{W}_\alpha) = R(\tilde{Q}) = R(\tilde{W}_\alpha) = 0$ (with $\tilde{Q}$ being the matter fields and $\tilde{W}_\alpha$ the gauge fields, but not in the canonically normalized form), then we may write
\begin{align}
\mathcal{W} 	& = \widetilde{\mathcal{W}} X_{\mathcal{W}}	&
\mathcal{K}	& = \widetilde{\mathcal{K}} X_{\mathcal{K}}
\end{align}
where the ``tilded" potentials are functions of only the ``tilded" fields.  Since the ``tilded" fields have no charges, the resulting potentials don't either; hence all the transformational weights belong to the $X_n$:
\begin{align*}
d_W(X_{\mathcal{K}})	& = +2		& d_W(X_{\mathcal{W}})		& = +3		\\
R(X_{\mathcal{K}})	& = 0		& R(X_{\mathcal{W}})		& = +2
\end{align*}

Now because the $X_n$ carry charges, they can only depend on the conformal compensator $\phi$ (we've already removed any other fields' dependence into the potentials).  Therefore invariance necessitates
\begin{align}
X_{\mathcal{K}} 	& = \phi^\dagger \phi		&
X_{\mathcal{W}}	& = \phi^3
\end{align}

We can now write the most general superconformal invariant lagrangian.  It is given by
\begin{equation}
\label{Eq:AMSB:gen.inv.L}
\mathcal{L}
	= \half \intOp[4]{\theta} \phi^\dagger \phi \widetilde{\mathcal{K}} \inp{\tilde{D}_\alpha, \tilde{Q}, \tilde{W}_\alpha} 
	  + \intOp[2]{\theta} \phi^3 \widetilde{\mathcal{W}} \inp{\tilde{Q}, \tilde{W}_\alpha}
	  + \text{h.c.}
\end{equation}

This picture explicitly demonstrates the $\phi$ couplings as required by superconformal invariance at a cost of using non-canonically normalized fields.  It is possible to return to the usual fields by defining
\begin{align}
\label{Eq:AMSB:field.redef}
Q	& = \phi \tilde{Q}					&
D_\alpha	& = \frac{\phi^\dagger}{\phi^{1/2}} \tilde{D}_\alpha	&
W_\alpha	& = \phi^{3/2} \tilde{W}_\alpha
\end{align}
with the last equation being a consequence of the second.  To illustrate how these definitions return us to the canonical fields, we must write the potentials schematically as
\begin{align}
\widetilde{\mathcal{K}}
	& =	\tilde{Q}^\dagger \EE^W \tilde{Q} + \ldots		\\
\widetilde{\mathcal{W}}
	& =	L \tilde{Q} + M \tilde{Q}^2 + Y \tilde{Q}^3 + \frac{\lambda}{\Lambda} \tilde{Q}^4 + \ldots + \tilde{W}^\alpha \tilde{W}_\alpha + \ldots
\end{align}
It is then clear that the lagrangian of \eq{AMSB:gen.inv.L}, combined with the field redefinitions \eq{AMSB:field.redef}, leads to a lagrangian
\begin{align}
\mathcal{L} 
	& = \half \intOp[4]{\theta} \inb{Q^\dagger \EE^W Q + \ldots}
\notag	\\
	& \quad {}
	  + \intOp[2]{\theta} \inb{L \phi^2 Q + M \phi Q^2 + Y Q^3 + \frac{\lambda}{\Lambda \phi} Q^4 + \ldots + W^\alpha W_\alpha + \ldots}
\notag	\\
	& \quad {}
	  + \text{h.c.}
\label{Eq:AMSB.gen.canon.L}
\end{align}

Several comments are in order regarding \eq{AMSB.gen.canon.L}: first, the presence of a linear or a mass term leave a $\phi$ in the superpotential resulting in the breaking of superconformal invariance at tree level---something relevant for the MSSM because of the Higgs's mass term.  Second, the nonrenormalizable terms always break superconformal invariance, and will always have the pair $\Lambda \phi$ to some power---as these terms are usually thought of as the result of a threshold or cutoff, this form will be important when we discuss intermediate thresholds and renormalization below.  Finally, if no dimensionful couplings are included ($L \rightarrow 0$, $M \rightarrow 0$, $\Lambda \rightarrow \infty$), the lagrangian is superconformal invariant at tree level; however, this is not true at loop level.

When including quantum corrections a mass parameter, $\mu$, will be introduced upon which the couplings (and the wave function renormalization constant $Z$) depend.  The mass parameter will also require some type of regulator which can be chosen to be a cutoff $\Lambda$.  This regulator is convenient to use because we have already established that such a cutoff must be paired with $\phi$ should it give rise to nonrenormalizable terms of the form in \eq{AMSB.gen.canon.L}\footnote{The result that the UV cutoff gets paired with a $\phi$ is independent of whether or not it yields nonrenormalizable terms; however, it is a convenient illustration here}.  Thus, because it is necessary for $\mu$ to always appear in the ratio $\mu/\abs{\Lambda}$, the effect of $\mu$ is to promote the renormalized parameters to superfields through the rule 
\begin{equation}
\mu \rightarrow \frac{\mu}{ \sqrt{\phi^\dagger \phi} }
\end{equation}

The promotion of $Z(\mu)$ to a superfield $\mathcal{Z}(\mu)$ and $1/g^2(\mu)$ to the superfield $\mathcal{R}(\mu)$ gives rise to soft SUSY breaking terms.  To obtain an expression for those terms it is convenient to chose a gauge where
\begin{equation}
\phi = 1 + F_\phi \theta^2
\end{equation}
This leads to the following form for the soft SUSY breaking parameters
\begin{align}
\notag
m_Q^2	& =	\half \abs{F_\phi}^2 \deriv{}{\ln \mu} \gamma_Q\inp[.5cm]{Y(\ln \mu), g_b(\ln \mu)}
	\\
\label{Eq:AMSB.scalar.mass}
	& = 	\half \abs{F_\phi}^2 \inb{	  \half \beta_{g_a} \pderiv{\gamma_Q}{g_a} 
						+ \beta_Y \pderiv{\gamma_Q}{Y} 
						+ \text{c.c.}
					}
	\\
\label{Eq:AMSB.trilinear.A}
A_Q	& =	\beta_{Y_Q} F_\phi
	\\
\label{Eq:AMSB.gaugino.mass}
M_{\lambda_a}
	& =	\frac{\beta_{g_a}}{g_a} F_\phi
\end{align}

For the MSSM the contribution of the second term in \eq{AMSB.scalar.mass} is negligible for at least the first and second generation sleptons, so the first term dominates.  As both $SU(2)_L\times U(1)_Y$ are infrared free, their $\beta$ functions are negative and hence the sleptons get negative mass-squares.  This is a fundamental problem because it implies the breakdown of electric charge in the ground state.  Before AMSB models can be phenomenologically viable this problem must be solved, but it is worth pursuing a solution because AMSB provides decoupling of UV physics in an elegant manner (we discuss this below), naturally suppressed FCNC (the SUSY breaking parameters depend on the Yukawa couplings and are diagonalized with them), and high predictive power with a minimal number of arbitrary parameters (essentially all soft SUSY breaking terms depend on only $F_\phi$).  It is therefore of great interest to seek reasonable models where the slepton mass-squares are made positive without destroying those good features.   We will present such a model in \Sec{SUSYLR}, where we demonstrate that extending the MSSM to include neutrino mass---generated by an $R$-parity conserving seesaw mechanism---will simply and effectively achieve this goal.  Yet for the moment we will continue our review of AMSB and address the decoupling of higher scale physics.

To illustrate the UV insensitivity of AMSB, consider a threshold $\Lambda \gg M \gg F_\phi$---such a scale may be an explicit mass term in the superpotential or the vev of the scalar component of the superfield $X$.  In either case we assume that below $M$ there are no remnant singlets in the effective theory; this is the same as requiring that as $\Lambda \rightarrow \infty$, $M$ remains finite.  The previous condition ensures that the effective theory's lagrangian has the schematic form\footnote{we use the notation $\scalar{X}$ as the scalar component of the superfield $X$}
\begin{align}
\notag
\mathcal{L}_{\text{eff}} 
	& =	\mathcal{L}_Q + M^{-n} f\inp{\scalar{Q},\psi_{Q}} + M^4
	\\
\label{Eq:AMSB.threshold.eff.L}
	& =	\half \mathcal{L}_Q + M^4 + \intOp[2]{\theta} \inp{\frac{Q^4}{M \phi} + \frac{Q^5}{M^2 \phi^2} + \ldots} + \text{h.c.}
\end{align}
where $n > 0$ and $\mathcal{L}_Q$ represents the part of the lagrangian involving only the various components of the matter superfields $Q$.  This form of the effective theory (which is of the exact same form as the nonrenormalizable terms of \eq{AMSB.gen.canon.L}) makes explicit that the additional SUSY breaking effects from the threshold $M$ go as $F_\phi / M \ll 1$.  Thus, the rule $\mu \rightarrow \mu/\abs{\phi}$ completely parameterizes all the SUSY breaking in both the high-scale and low-scale theories resulting in the maintenance of the AMSB trajectory below $M$.

Another means to see this decoupling is to realize that the replacement rule due to the threshold is 
\begin{equation}
\label{Eq:AMSB.threshold.rule}
M \rightarrow M \phi + F_\phi \inb{ c_1 \frac{F_\phi}{M} + c_2 \pfrac{F_\phi}{M}^2 + \cdots }\theta^2 \approx M \phi
\end{equation}
in addition to which there is the requirement pairing $\Lambda$ with $\phi$.  The quantum corrections of the lagrangian of \eq{AMSB.threshold.eff.L} force $M$ to appear in the effective theory as $\ln \abs{M/\Lambda}$ and  $\ln \mu/\abs{M}$ (which comes when $M$ is used as a cutoff in loop calculations).  Using the replacement rules on these quantities gives
\begin{align}
\ln \abs{\frac{M}{\Lambda}} 	
	& \rightarrow \ln \abs{\frac{M \phi}{\Lambda \phi}} = \ln \abs{\frac{M}{\Lambda}}
	\\
\ln \frac{\mu}{\abs{M}}
	& \rightarrow \ln \frac{\mu}{\abs{M \phi}}
\end{align}
and once again only $\mu \rightarrow \mu/\abs{\phi}$ is required to capture all the SUSY breaking.

The above argument may disturb the reader since the $\beta$ functions change when crossing the threshold; however, what is actually happening is that when the threshold is crossed, the removal of the heavy fields adds a term $\Delta \beta$ that results in a shift of the higher-scale $\beta$ functions, $\beta^+$, to the lower-scale $\beta$ functions, $\beta^-$.   This property, namely
\begin{equation}
\Delta \beta + \beta^+ = \beta^-
\end{equation}
is the one that keeps the theory in the AMSB form.

The UV decoupling of AMSB presents a major obstacle for fixing the negative mass-squares of the MSSM since any high-scale tinkering will leave little to no evidence at the low scale.

\ifglobal{}%
\else%
%
\bibliography{\bibpath/susy,\bibpath/amsb,\bibpath/susy-ext_to_superspace}

\end{document}
\fi%

\section{SUSYLR and AMSB}
\label{Sec:SUSYLR}

\makeatletter%
\@ifundefined{ifglobal}%
{%
\newif\ifglobal%
\globalfalse%
\newif\ifprd%
\prdfalse%
}%
\makeatother%
\ifglobal{}%
\else%
\documentclass[12pt]{article}

\usepackage{ifthen}
\usepackage{calc}
\ifprd{}
\else
	\usepackage{amsmath}
	\usepackage{amssymb}
	\usepackage{url}
\fi
\usepackage{graphicx}	
\usepackage{mathrsfs}	
\usepackage{bbm}		
\usepackage{natbib}	
\usepackage[usenames]{color}	

\newcommand{\ourtitle}{Minimal Seesaw as an Ultraviolet Insensitive Cure for the Problems of Anomaly Mediation}
\newcommand{\paperauth}{R. N. Mohapatra, N. Setzer, and S. Spinner}
\newcommand{\ouraffiliation}{Maryland Center for Fundamental Physics and Department of Physics, University of Maryland, College Park, MD 20742, USA}
\newcommand{\ourpreprint}{\vbox{ \hbox{UMD-PP-XX-XXX} }}
\newcommand{\ourpacs}{14.60.Pq, 98.80.Cq}
\newcommand{\ourdate}{}

\ifprd{}
\else
\title{\ourtitle}
\author{\paperauth}	
\fi


\newcommand{\eq}[1]{Eq.~\eqref{Eq:#1}}
\newcommand{\eqn}[1]{\eqref{Eq:#1}}
\newcommand{\fig}[1]{Figure~\ref{Fig:#1}}
\newcommand{\Sec}[1]{Section~\ref{Sec:#1}}
\newcommand{\tbl}[1]{Table~\ref{Table:#1}}
\newcommand{\app}[1]{Appendix~\ref{App:#1}}
\newcounter{itemnum}
\newenvironment{olist}
{%
\begin{list}%
	{(\roman{itemnum})}%
	{
	\usecounter{itemnum}}%
}%
{%
\end{list}
}%

\let\gobbleEnd=\relax
\newcommand{\drawmasslines}[2]{\gobblemasslist#2,\gobbleHalt\gobbleEnd}
\newcommand{\gobbleHalt}{\@\@\@\@}

\def\gobblemasslist(#1,#2,[#3,#4]),#5\gobbleEnd{%
\ifthenelse{ \equal{#5}{\gobbleHalt} }%
	{%
	\put(\xCoord,#2){\line(1,0){\masslinelen} \raisebox{#4}{\hspace{#3}$#1$}}%
	}%
	{%
	\put(\xCoord,#2){\line(1,0){\masslinelen} \raisebox{#4}{\hspace{#3}$#1$}}%
	\gobblemasslist#5\gobbleEnd%
	}%
}%

\newlength{\oldunitlength}
\setlength{\oldunitlength}{\the\unitlength}
\newcommand{\masslinelen}{50}
\newcommand{\xCoordColOne}{10}
\newcommand{\xCoordColTwo}{120}
\newcommand{\xCoordColThree}{230}
\newcommand{\xCoordColFour}{340}
\newcommand{\xCoordColFive}{450}
\newcommand{\xCoord}{}
\newcommand{\defaultX}{1mm}
\newcommand{\defaultY}{-0.6ex}
\newcommand{\placetoleft}{-1.4cm}
\newcommand{\placetoleftnudgeright}{-1.33cm}
\newcommand{\slightlyraised}{-0.4ex}
\newcommand{\slightlylowered}{-0.8ex}


\newcommand{\E}[1]{\times 10^{#1}}	
\newcommand{\vev}[1]{\inap{#1}}		

\newcommand{\MP}{M_{\text{Pl}}} 		
\newcommand{\MSUSY}{M_\text{SUSY}}	
\newcommand{\Mmess}{M_\text{mess}}	

\newcommand{\G}[1]{\ensuremath{G_{#1}}}	
\newcommand{\SUSYLR}{SUSYLR}
\newcommand{\CISLR}{CISLR}

\newcommand{\WSUSYLR}{W_{\text{\SUSYLR}}}

\newcommand{\ourmodel}{AMSBLR}

\newcommand{\DeltaBar}{\bar{\Delta}}

\newcommand{\DeltaC}{\Delta^c}
\newcommand{\DeltaBarC}{\bar{\Delta}^c}

\newcommand{\scalar}[1]{%
\ifthenelse{ \equal{#1}{\DeltaC} }%
	{%
	\underline{\Delta}^c%
	}%
{ \ifthenelse{ \equal{#1}{\DeltaBarC} }%
	{%
	\underline{\bar{\Delta}}^c%
	}%
{ \ifthenelse{ \equal{#1}{\Phi} }%
	{%
	\underline{\Phi}%
	}%
	{%
	\underline{#1}%
	}}}%
}%

\newcommand{\DBarCpp}{\bar{\Delta}^{c ++}}
\newcommand{\DBarCp}{\bar{\Delta}^{c +}}
\newcommand{\DBarCz}{\bar{\Delta}^{c \,0}}
\newcommand{\DCmm}{\Delta^{c --}}
\newcommand{\DCm}{\Delta^{c -}}
\newcommand{\DCz}{\Delta^{c \,0}}
\newcommand{\phiUp}{\Phi_u^+}
\newcommand{\phiUz}{\Phi_u^0}
\newcommand{\phiDz}{\Phi_d^0}
\newcommand{\phiDm}{\Phi_d^-}

\newcommand{\Slight}{N}
\newcommand{\Sheavy}{S}

\newcommand{\kappaLight}{\kappa_\Slight}
\newcommand{\kappaHeavy}{\kappa_\Sheavy}

\newcommand{\lambdaLight}{\lambda_\Slight}
\newcommand{\lambdaHeavy}{\lambda_\Sheavy}

\newcommand{\I}{\mathbbm{i}}
\newcommand{\EE}{\mathbbm{e}}

\newcommand{\bibpath}{.}
\bibliographystyle{h-physrev}
\citestyle{plain}

\begin{document}

\maketitle

\section{SUSYLR and AMSB}
\label{Sec:SUSYLR}

\fi%

The new feature of models combining AMSB and \SUSYLR{} is that the effective theory below the $v_R$ scale contains Yukawa couplings to both the left- and right-handed electrons in addition to those of the MSSM; hence the slepton masses can be made positive.  Thus, the marriage of \SUSYLR{} with AMSB gives positive slepton mass-squares and the resulting theory combines the prodigious predictive power of AMSB, explains small neutrino masses (through the seesaw mechanism), and retains a natural dark matter candidate (the LSP is stable due to $R$-parity conservation).

\subsection{The Model}

The particle content of a \SUSYLR{} model is shown in \tbl{QM Numbers}.  As the model is left-right symmetric, it contains both left- and right-handed higgs bosons---in this case $B - L = \pm 2$ triplets so that $R$-parity may be preserved (a task for which the $B - L = 1$ doublets are not suitable).  The presence of $SU(2)_L$ and $SU(2)_R$ triplets means that parity is a good symmetry until $SU(2)_R$ breaks.  While the seesaw mechanism may be achieved with only $SU(2)_R$ higgs fields, demanding parity forces the left-handed triplets to be present these together then yield positive slepton masses.

\begin{table}[ht]
\begin{center}
\begin{tabular}{|c|c|}
\hline\hline
Fields			& $SU(3)^c	\times	SU(2)_L	\times	SU(2)_R	\times	U(1)_{B-L}$		\\
\hline
$Q$			& $(3		,	2	,	1	,	+\frac{1}{3})$		\\
$Q^c$			& $(\bar{3}	,	1	,	2	,	-\frac{1}{3})$		\\
$L$			& $(1		,	2	,	1	,	-1)$			\\
$L^c$			& $(1		,	1	,	2	,	+1)$			\\
$\Phi$			& $(1		,	2	,	2	,	0)$			\\
$\Delta$			& $(1		,	3	,	1	,	+2)$			\\
$\DeltaBar$		& $(1		,	3	,	1	,	-2)$			\\
$\DeltaC$		& $(1		,	1	,	3	,	-2)$			\\
$\DeltaBarC$		& $(1		,	1	,	3	,	+2)$			\\
\hline\hline
\end{tabular}
\end{center}
\caption{Assignment of the fermion and Higgs fields' representations of the left-right symmetry group (except for $U(1)_{B-L}$ where the charge under that group is given.)}
\label{Table:QM Numbers}
\end{table}

The parity-respecting \SUSYLR{} superpotential is then
\begin{align}
\label{Eq:SuperW.SUSYLR}
\WSUSYLR
	& = W_{\text{Y}} + W_{\text{H}} + W_{\text{NR}}
\end{align}
with
\begin{align}
W_{\text{Y}}
	& = 	  \I y_{Q}^a Q^T \tau_2 \Phi_a Q^c 
		+ \I y_{L}^a L^T \tau_2 \Phi_a L^c
		+ \I f_c L^{cT} \tau_2 \DeltaC L^c
		+ \I f L^T \tau_2 \Delta L
\label{Eq:SuperW.SUSYLR.Yuk}
	\\
W_{\text{H}}
	& =	   \inp{ M_{\Delta} \phi - \lambdaHeavy \Sheavy } \inb{ \Tr\inp{\DeltaC \DeltaBarC} + \Tr\inp{\Delta \DeltaBar} }
		+ \half \mu_{\Sheavy} \phi \Sheavy^2
		+ \third \kappaHeavy \Sheavy^3
\notag	\\
	& \quad {}
		+ \lambdaLight^{ab} \Slight \Tr\inp{\Phi_a^T \tau_2 \Phi_b \tau_2}
		+ \third \kappaLight \Slight^3
\label{Eq:SuperW.SUSYLR.Higgs}
	\\
W_{\text{NR}}
	& =	  \frac{\lambda_A}{\MP \phi} \Tr^2\inp{\Delta \DeltaBar}
		+ \frac{\lambda_A^c}{\MP \phi} \Tr^2\inp{\DeltaC \DeltaBarC}
\notag	\\
	& \quad {}
		+ \frac{\lambda_B}{\MP \phi} \Tr\inp{\Delta \Delta} \Tr\inp{\DeltaBar \DeltaBar}
		+ \frac{\lambda_B^c}{\MP \phi} \Tr\inp{\DeltaC \DeltaC} \Tr\inp{\DeltaBarC \DeltaBarC}
\notag	\\
	& \quad {}
		+ \frac{\lambda_C}{\MP \phi} \Tr\inp{\Delta \DeltaBar} \Tr\inp{\DeltaC \DeltaBarC}
\notag	\\
	& \quad {}
		+ \frac{\lambda_D}{\MP \phi} \Tr\inp{\Delta \Delta} \Tr\inp{\DeltaC \DeltaC}
		+ \frac{\bar\lambda_D}{\MP \phi} \Tr\inp{\DeltaBar \DeltaBar} \Tr\inp{\DeltaBarC \DeltaBarC}
		+ \cdots
\label{Eq:SuperW.SUSYLR.nr}
\end{align}
We have assumed that the singlet couplings absent from \eq{SuperW.SUSYLR.Higgs} are zero or small enough that they can be neglected.  This condition is necessary to keep one singlet light ($\Slight$) so that below the right-handed scale $v_R$ the theory is the NMSSM with some additional particles.  Although this may seem rather \textit{ad hoc}, we do it out of convenience rather than necessity: the low scale theory must be such that it avoids an MSSM higgs bilinear $b$ term that is too large\cite{Luty:2005sn}; the superpotential given above happens to be one.  However, it is not the \emph{only} one and several alternative methods exist\cite{Katz:1999uw,Randall:1998uk} to avoid this problem.  As any of these alternatives are equally valid, and because the exact form of the electroweak scale theory is irrelevant to the conclusions, we merely select to use the superpotential above.

The superpotential of \eq{SuperW.SUSYLR.Higgs} dictates that
\begin{align}
\label{Eq:vev.S.heavy}
\vev{\Sheavy}
	& = 	\frac{ M_{\Delta} }{ \lambdaHeavy } \phi		\\
\label{Eq:vev.Dc.Dbarc}
\vev{\DeltaC} \vev{\DeltaBarC}
	& = \vev{\Sheavy}\inp{	   \frac{ M_{\Delta} \kappaHeavy }{ \lambdaHeavy^2 }
				 + \frac{ \mu_{\Sheavy} }{ \lambdaHeavy }
				} \phi
\end{align}
\eq{vev.S.heavy} should be evident from the form of the superpotential; \eq{vev.Dc.Dbarc} requires \eq{SuperW.SUSYLR.Higgs} to be recast as
\begin{equation}
W_{\text{H}}
	\supset		\inb{
				- \lambdaHeavy  \Tr\inp{\DeltaC \DeltaBarC}
				+ \half \mu_{\Sheavy} \phi \Sheavy
				+ \third \kappaHeavy \Sheavy^2
			} \Sheavy
\end{equation}
The inclusion of the nonrenormalizable terms of \eq{SuperW.SUSYLR.nr} (which are necessary if $R$-parity is conserved \cite{Aulakh:1998nn,Chacko:1997cm})will shift the vevs\footnote{We denote the scalar component of the superfield $X$ as $\scalar{X}$} of $\scalar{\DeltaC}$, $\scalar{\DeltaBarC}$, and $\scalar{\Sheavy}$ by $\sim M_{\Delta}^2 / \MP \ll M_{\Delta}$ so they may be safely be ignored.  It is worth noting that the nonrenormalizable terms are only irrelevant because as $\MP \rightarrow \infty$ the vevs all remain finite; that is, they depend at most on $1/\MP$.

Because the nonrenormalizable operators are insignificant, Eqs.~\eqn{vev.S.heavy} and \eqn{vev.Dc.Dbarc} are still valid and the theory respects the AMSB trajectory below $v_R$: the advertised UV insensitivity.  Yet even though the particles remain on their AMSB trajectory, the negative slepton mass-squares problem is still solved.  This comes about because of the additional yukawa couplings $f$ and $f_c$ which survive to the lower-scale theory.

The existence of the $f$ coupling at the lower scale can be seen from the superpotential \eq{SuperW.SUSYLR.Higgs}: when $S$ gets the vev of \eq{vev.S.heavy}, the mass term for the $SU(2)_L$ triplets vanishes while the $SU(2)_R$ triplets also get a vev, so their mass term remains.  This would leave $\Delta$ and $\DeltaBar$ massless below the right-handed breaking scale except that the non-renormalizable terms contribute a mass through
\begin{equation}
\frac{\lambda_C}{\MP \phi} \Tr\inp{\Delta \DeltaBar} \Tr\inp{\DeltaC \DeltaBarC}
	\rightarrow	\frac{\lambda_C}{\MP \phi} \vev{\DeltaC} \vev{\DeltaBarC} \Tr\inp{\Delta \DeltaBar}
	\simeq		\frac{\lambda_C v_R^2 \phi}{\MP} \Tr\inp{\Delta \DeltaBar}
\end{equation}

The same mass value of $v_R^2/\MP$ is also responsible for $f_c$ surviving to the low scale, but this time in the context of light-doubly charged particles.  It is well known that the class of \SUSYLR{} models considered here have light doubly-charged particles\cite{Chacko:1997cm} with a mass as mentioned above.  The question that needs to be addressed here is ``how light?"  If their mass is large, $F_\phi \ll m_{DC} \ll v_R$, then these particles merely introduce another trajectory preserving threshold which decouples from the lower scale theory.  For the right-handed selectron this would be disastrous as it would have a purely negative AMSB contribution to its mass.  Thus, it makes sense to demand that the doubly-charged particles have a mass $m_{DC} \lesssim F_\phi$.

The existence of the $SU(2)_L$ triplets and the doubly-charged particles below or around $m_{3/2}$ means that their couplings remain in the low-scale superpotential and are therefore important.  For the sleptons, the relevant terms are
\begin{equation}
W \supset f_c \DCmm e^c e^c + \I f L^T \tau_2 \Delta L
\end{equation}

The survival of these yukawa couplings $f_c$ and $f$ allow the scalar $e^c$ and $e$ mass-squares to be positive.  Assuming that $f$, $f_c$ are diagonal in flavor space (an assumption validated by lepton flavor violating experiments\cite{Bellgardt:1987du}), we need only $f_1 \simeq f_2 \simeq \nop{f_{c}}_1 \simeq \nop{f_{c}}_2 \simeq \mathcal{O}(1)$ to make the sleptons positive.  The only constraint here is from muonium-antimuonium oscillations\cite{Willmann:1998gd} which demands that $\nop{f_{c}}_1 \nop{f_{c}}_2 / 4 \sqrt{2} m_{DC}^2 \approx f_1 f_2 / 4 \sqrt{2} m_{DC}^2 < 3 \E{-3} G_F$; however, with both the doubly-charged fields and $SU(2)_L$ triplets having a mass $m_{DC} \simeq F_\phi \sim 10$ TeV, this is easily satisfied.  Furthermore, this constraint limits the range for $v_R$ as $m_{DC} \simeq v_R^2 /\MP \simeq F_\phi$ implies that $v_R \simeq 10^{11}$--$10^{12}$ GeV.

The amazing result is that AMSB and \SUSYLR{} yield a sfermion sector that depends on very few parameters: $F_\phi$, $\nop{f_c}_1$, $\nop{f_c}_3$, in addition to the usual $\tan \beta$ and $\sgn{\mu}$ (because of parity, $f_1 = \nop{f_c}_1$ and $f_3 = \nop{f_c}_3$).  Interestingly, two of the new parameters---the $f_c$ yukawa couplings---also have implications for neutrino oscillations.

\subsection{Numerical Analysis}

We now present the resulting mass spectrum for this model.  For this analysis we start by running the parameters of the Standard Model up to $\MSUSY{}$, match at that point to the NQNMSSM (Not-Quite NMSSM: the NMSSM with doubly-charged particles, left-handed triplets, and two additional Higgs doublets), and use the appropriately modified RGEs of \cite{Setzer:2005hg} to get to the right-handed scale.  Without loss of generality we assume that only one up-type Higgs and one down-type Higgs get a vev\cite{Drees:1988fc}.  Additionally, we take the standard simplifying assumption that only the third generation higgs yukawa couplings are important.

\begin{figure}
\begin{center}
\includegraphics[scale=0.8]{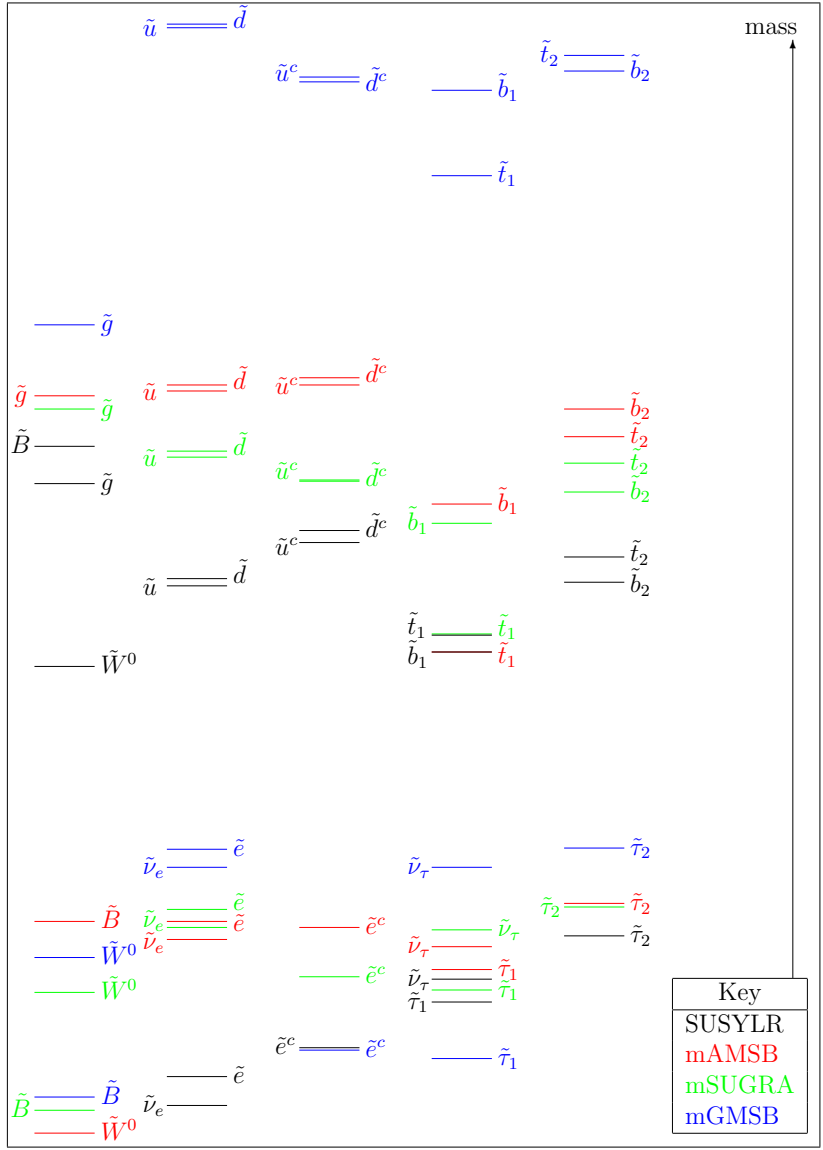}
\end{center}
\caption{%
The mass spectrum for the superpartners of the Standard Model for four different models (in four different colors):
\SUSYLR$+$AMSB, \textcolor{red}{mAMSB}, \textcolor{green}{mSUGRA}, and \textcolor{blue}{mGMSB}.  Note that for the \SUSYLR$+$AMSB, $\tilde{t}_2$ and $\tilde{b}_2$ are mostly right-handed; in contrast with the usual mSUGRA or mGMSB cases where they are typically mostly left-handed.%
}
\label{Fig:SuperPartner.Masses}
\end{figure}

\fig{SuperPartner.Masses} shows the mass spectrum of the general \SUSYLR{} model and the MSSM with other popular SUSY breaking scenarios (the figure is truly only comprehensible in color---a form available on line at \url{http://arXiv.org}).  The comparison was obtained by matching the gluino mass between the models, and then running the masses down to the scale $Q$ using ISAJET\cite{Paige:2003mg}.  The spectra in \fig{SuperPartner.Masses} contain the generic features, though the figure was generated using the points listed in \tbl{benchmark.pts}.  It is also interesting to note that the heavier sfermion mass eigenstates are mostly right-handed contrary to most mSUGRA and GMSB scenarios.

\begin{table}
\begin{center}
\begin{tabular}{|c|c|c|c|}
\hline\hline
\SUSYLR$+$AMSB		& AMSB$+m_0^2$	 	& mGMSB 			& mSUGRA 		\\
\hline
$\tan \beta = 15$	& $\tan \beta = 15$	& $\tan \beta = 15$	& $\tan \beta = 15$	\\
$\sgn \mu = +1$		& $\sgn \mu = +1$	& $\sgn \mu = +1$	& $\sgn \mu = +1$	\\
$Q = 550$ GeV		& $Q = 558$ GeV		& $Q = 899$ GeV		& $Q = 537$ GeV		\\
$F_\phi = 30$ TeV	& $F_\phi = 30$ TeV	& $\Lambda = 90$ TeV	& $m_0 = 190$ GeV	\\
			& $m_0 = 290$ GeV	& $\Mmess = 180$ TeV	& $m_{1/2} = 285$ GeV	\\
$v_R = 135$ EeV		&			&			& $A_{0} = 241$ GeV	\\
$f_1 = \nop{f_c}_1 = 0.52$
			&			&			&			\\
$f_3 = \nop{f_c}_3 = 0.6$
			&			&			&			\\

\hline\hline
\end{tabular}
\end{center}
\caption{The benchmark points for creating the spectrum of \fig{SuperPartner.Masses}.  The parameters shown were chosen by matching the gluino mass for \SUSYLR$+$AMSB to mGMSB; a Polonyi-like model was used for mSUGRA matching $m_0$ to $F_\phi/16 \pi^2$.  $Q$ is the scale at which the the masses are reported by ISAJET.  Because it is not widely known, we remind the reader that the metric prefix $E$ in the above table means ``exa" and is $10^{18}$.}
\label{Table:benchmark.pts}
\end{table}

One of the more striking features of the \SUSYLR+AMSB spectrum is that gaugino sector masses are all relatively close to each other.  This is unique from the popular scenarios displayed in \fig{SuperPartner.Masses} and is due to the contributions of the $SU(2)_L$ and $U(1)_Y$ extended particle content at low energy.  Such a massive wino consequently relaxes the naturalness arguments made in \cite{Feng:1999hg, Feng:1999fu}.  These arguments proceed along the lines that squark masses and the $\mu$ term must be below around $1$ TeV to preserve the naturalness of SUSY.  Therefore, a naturalness upper bound can be put on the wino mass.  Such an upper bound suggests that run II of the tevatron should have explored most of the viable wino parameter space, which would not be the case here.

Furthermore, we can achieve regions in parameter space where $F_\phi$ is lower than would be possible in other AMSB models without violating these naturalness bounds.  Specifically we can investigate a point in parameter space such as $F_\phi = 15$ TeV, $\tan \beta = 15$, $f_{c1} = f_1 = 1$ and $f_{c3} = f_3 = 1.6$ ($f$s are at the right-handed scale) with a spectrum given in \tbl{SUSYLR.TightMasses}.  Here even the sfermion sector has very little hierarchy in it.  Such spectra are exotic compared to typical mSUGRA and mGMSB type models although they are possible in deflected AMSB\cite{Rattazzi:1999qg}.

\begin{table}[ht]
\begin{center}
\begin{tabular}{|cc|}
\hline\hline
particle			& masses (GeV)	\\
\hline
$\tilde{t}_1$		& 291		\\
$\tilde{b}_1$		& 244		\\
$\tilde{u}$		& 296		\\
$\tilde{d}$		& 305		\\
$\tilde{t}_2$		& 348		\\
$\tilde{b}_2$		& 317		\\
$\tilde{u}^c$		& 314		\\
$\tilde{d}^c$		& 320		\\
$\tilde{\nu}_\tau$	& 195		\\
$\tilde{\tau}_1$		& 174		\\
$\tilde{\tau}_2$		& 236		\\
$\tilde{\nu}_e$		& 150		\\
$\tilde{e}$		& 169		\\
$\tilde{e}^c$		& 158		\\
$\tilde{B}$		& 326		\\
$\tilde{W}$		& 241		\\
$\tilde{G}$		& 340		\\
\hline\hline
\end{tabular}
\end{center}
\caption{Mass spectrum for the point $F_\phi = 15$ TeV, $\tan \beta = 15$ and at the right-handed scale $f_{c1} = f_1 = 1$ and $f_{c3} = f_3 = 1.6$.  Masses are evaluated at $Q = 325$ GeV.}
\label{Table:SUSYLR.TightMasses}
\end{table}

From a cosmological point of view, there is a potential problem with the increase in $SU(2)_L$ and $U(1)_Y$ gauge coupling strengths at the right-handed scale: they cause tachyonic squark masses at that scale (remember these gauge couplings give a negative contribution in \eq{AMSB.scalar.mass}).  Theories with tachyonic squark masses have been studied in the GUT framework and were found to be safe albeit unsavory\cite{Dermisek:2006ey}.  Large reheating temperatures will cause charge violating vacua to disappear\cite{Falk:1996zt} and tunneling rates to the bad vacua are too small in most of the parameter space\cite{Riotto:1995am, Kusenko:1996jn} to cause a problem.

Continuing along cosmological lines, both mass spectrums shown above indicate that the sneutrino is the LSP in this model.  Both the tau and electron sneutrinos are LSP candidates depending on the relative sizes of $f_3$ and $f_1$.  Although sneutrino dark matter is highly constrained\cite{Hall:1997ah, smith:2001hy}, there could be other dark matter candidates such as light singlet fields mixed with Higgsinos. It could also be that the sneutrinos generated from late decay of the gravitino are dark matter. We are currently investigating these scenarios.

Finally, let us consider the sleptons masses---the main purpose of this paper.  As advertised earlier, these are positive and depend on just a few parameters: $F_{\phi}, f_1, f_3$ (since we have preserve parity at the high scale in this paper $f_{c1} = f_1$ and $f_{c3} = f_3$ at the right-handed scale) and to a lesser extent on $\tan \beta$ and the right-handed scale.  The relative sizes of the masses are controlled by relative $f$ coupling: the larger the coupling the larger the mass, \textit{e.g.} increasing $f_1$ would raise the mass of the left-handed slepton.  Such an affect contrasts strongly with other non-AMSB models with light doubly-charged higgses where the right-handed stau mass drops with increase in $f_{c3}$ type coupling \cite{Dutta:1998bn, Setzer:2006sf}.

\ifglobal{}%
\else%
%
\bibliography{\bibpath/amsb,\bibpath/susy,\bibpath/susylr,\bibpath/flavor_violation,\bibpath/susy-extended_higgs,Pheno}

\end{document}
\fi%

\section{Conclusion}
\label{Sec:Conclusion}

\makeatletter%
\@ifundefined{ifglobal}%
{%
\newif\ifglobal%
\globalfalse%
\newif\ifprd%
\prdfalse%
}%
\makeatother%
\ifglobal{}%
\else%
\documentclass[12pt]{article}

\usepackage{ifthen}
\usepackage{calc}
\ifprd{}
\else
	\usepackage{amsmath}
	\usepackage{amssymb}
	\usepackage{url}
\fi
\usepackage{graphicx}	
\usepackage{mathrsfs}	
\usepackage{bbm}		
\usepackage{natbib}	
\usepackage[usenames]{color}	

\newcommand{\ourtitle}{Minimal Seesaw as an Ultraviolet Insensitive Cure for the Problems of Anomaly Mediation}
\newcommand{\paperauth}{R. N. Mohapatra, N. Setzer, and S. Spinner}
\newcommand{\ouraffiliation}{Maryland Center for Fundamental Physics and Department of Physics, University of Maryland, College Park, MD 20742, USA}
\newcommand{\ourpreprint}{\vbox{ \hbox{UMD-PP-XX-XXX} }}
\newcommand{\ourpacs}{14.60.Pq, 98.80.Cq}
\newcommand{\ourdate}{}

\ifprd{}
\else
\title{\ourtitle}
\author{\paperauth}	
\fi


\newcommand{\eq}[1]{Eq.~\eqref{Eq:#1}}
\newcommand{\eqn}[1]{\eqref{Eq:#1}}
\newcommand{\fig}[1]{Figure~\ref{Fig:#1}}
\newcommand{\Sec}[1]{Section~\ref{Sec:#1}}
\newcommand{\tbl}[1]{Table~\ref{Table:#1}}
\newcommand{\app}[1]{Appendix~\ref{App:#1}}
\newcounter{itemnum}
\newenvironment{olist}
{%
\begin{list}%
	{(\roman{itemnum})}%
	{
	\usecounter{itemnum}}%
}%
{%
\end{list}
}%

\let\gobbleEnd=\relax
\newcommand{\drawmasslines}[2]{\gobblemasslist#2,\gobbleHalt\gobbleEnd}
\newcommand{\gobbleHalt}{\@\@\@\@}

\def\gobblemasslist(#1,#2,[#3,#4]),#5\gobbleEnd{%
\ifthenelse{ \equal{#5}{\gobbleHalt} }%
	{%
	\put(\xCoord,#2){\line(1,0){\masslinelen} \raisebox{#4}{\hspace{#3}$#1$}}%
	}%
	{%
	\put(\xCoord,#2){\line(1,0){\masslinelen} \raisebox{#4}{\hspace{#3}$#1$}}%
	\gobblemasslist#5\gobbleEnd%
	}%
}%

\newlength{\oldunitlength}
\setlength{\oldunitlength}{\the\unitlength}
\newcommand{\masslinelen}{50}
\newcommand{\xCoordColOne}{10}
\newcommand{\xCoordColTwo}{120}
\newcommand{\xCoordColThree}{230}
\newcommand{\xCoordColFour}{340}
\newcommand{\xCoordColFive}{450}
\newcommand{\xCoord}{}
\newcommand{\defaultX}{1mm}
\newcommand{\defaultY}{-0.6ex}
\newcommand{\placetoleft}{-1.4cm}
\newcommand{\placetoleftnudgeright}{-1.33cm}
\newcommand{\slightlyraised}{-0.4ex}
\newcommand{\slightlylowered}{-0.8ex}


\newcommand{\E}[1]{\times 10^{#1}}	
\newcommand{\vev}[1]{\inap{#1}}		

\newcommand{\MP}{M_{\text{Pl}}} 		
\newcommand{\MSUSY}{M_\text{SUSY}}	
\newcommand{\Mmess}{M_\text{mess}}	

\newcommand{\G}[1]{\ensuremath{G_{#1}}}	
\newcommand{\SUSYLR}{SUSYLR}
\newcommand{\CISLR}{CISLR}

\newcommand{\WSUSYLR}{W_{\text{\SUSYLR}}}

\newcommand{\ourmodel}{AMSBLR}

\newcommand{\DeltaBar}{\bar{\Delta}}

\newcommand{\DeltaC}{\Delta^c}
\newcommand{\DeltaBarC}{\bar{\Delta}^c}

\newcommand{\scalar}[1]{%
\ifthenelse{ \equal{#1}{\DeltaC} }%
	{%
	\underline{\Delta}^c%
	}%
{ \ifthenelse{ \equal{#1}{\DeltaBarC} }%
	{%
	\underline{\bar{\Delta}}^c%
	}%
{ \ifthenelse{ \equal{#1}{\Phi} }%
	{%
	\underline{\Phi}%
	}%
	{%
	\underline{#1}%
	}}}%
}%

\newcommand{\DBarCpp}{\bar{\Delta}^{c ++}}
\newcommand{\DBarCp}{\bar{\Delta}^{c +}}
\newcommand{\DBarCz}{\bar{\Delta}^{c \,0}}
\newcommand{\DCmm}{\Delta^{c --}}
\newcommand{\DCm}{\Delta^{c -}}
\newcommand{\DCz}{\Delta^{c \,0}}
\newcommand{\phiUp}{\Phi_u^+}
\newcommand{\phiUz}{\Phi_u^0}
\newcommand{\phiDz}{\Phi_d^0}
\newcommand{\phiDm}{\Phi_d^-}

\newcommand{\Slight}{N}
\newcommand{\Sheavy}{S}

\newcommand{\kappaLight}{\kappa_\Slight}
\newcommand{\kappaHeavy}{\kappa_\Sheavy}

\newcommand{\lambdaLight}{\lambda_\Slight}
\newcommand{\lambdaHeavy}{\lambda_\Sheavy}

\newcommand{\I}{\mathbbm{i}}
\newcommand{\EE}{\mathbbm{e}}

\newcommand{\bibpath}{.}
\bibliographystyle{h-physrev}
\citestyle{plain}

\begin{document}

\maketitle

\section{Conclusion}
\label{Sec:Conclusion}

\fi%

We have presented a new way to solve the negative mass-squared slepton problem of AMSB using a minimal, bottom-up extension of the MSSM that incorporates neutrino masses (via the seesaw mechanism), solves the strong CP problem, and resolves the $R$-parity violation problem of the MSSM.  Slepton masses are rescued from the red by their couplings to both remnant doubly-charged fields and left-handed triplets.  Constraints from low energy physics and the non-decoupling of these additional fields require the seesaw scale to be around $10^{11}$ GeV clearly distinguishing our model from GUT seesaw models.

The model we presented has soft terms which remain on their AMSB trajectory down to the SUSY scale.  We have shown the sparticle spectrum for this model and compared it with typical predictions from other SUSY breaking scenarios finding significant deviations, especially in the gaugino sector.  Furthermore, in some regions of parameter space it is possible to produce a spectrum with little hierarchy between sleptons and squarks.

\ifglobal{}%
\else%
%
\bibliography{\bibpath/susy}

\end{document}
\fi%

\section{Acknowledgements}

\makeatletter%
\@ifundefined{ifglobal}%
{%
\newif\ifglobal%
\globalfalse%
\newif\ifprd%
\prdfalse%
}%
\makeatother%
\ifglobal{}%
\else%
\documentclass[12pt]{article}

\usepackage{ifthen}
\usepackage{calc}
\ifprd{}
\else
	\usepackage{amsmath}
	\usepackage{amssymb}
	\usepackage{url}
\fi
\usepackage{graphicx}	
\usepackage{mathrsfs}	
\usepackage{bbm}		
\usepackage{natbib}	
\usepackage[usenames]{color}	

\newcommand{\ourtitle}{Minimal Seesaw as an Ultraviolet Insensitive Cure for the Problems of Anomaly Mediation}
\newcommand{\paperauth}{R. N. Mohapatra, N. Setzer, and S. Spinner}
\newcommand{\ouraffiliation}{Maryland Center for Fundamental Physics and Department of Physics, University of Maryland, College Park, MD 20742, USA}
\newcommand{\ourpreprint}{\vbox{ \hbox{UMD-PP-XX-XXX} }}
\newcommand{\ourpacs}{14.60.Pq, 98.80.Cq}
\newcommand{\ourdate}{}

\ifprd{}
\else
\title{\ourtitle}
\author{\paperauth}	
\fi


\newcommand{\eq}[1]{Eq.~\eqref{Eq:#1}}
\newcommand{\eqn}[1]{\eqref{Eq:#1}}
\newcommand{\fig}[1]{Figure~\ref{Fig:#1}}
\newcommand{\Sec}[1]{Section~\ref{Sec:#1}}
\newcommand{\tbl}[1]{Table~\ref{Table:#1}}
\newcommand{\app}[1]{Appendix~\ref{App:#1}}
\newcounter{itemnum}
\newenvironment{olist}
{%
\begin{list}%
	{(\roman{itemnum})}%
	{
	\usecounter{itemnum}}%
}%
{%
\end{list}
}%

\let\gobbleEnd=\relax
\newcommand{\drawmasslines}[2]{\gobblemasslist#2,\gobbleHalt\gobbleEnd}
\newcommand{\gobbleHalt}{\@\@\@\@}

\def\gobblemasslist(#1,#2,[#3,#4]),#5\gobbleEnd{%
\ifthenelse{ \equal{#5}{\gobbleHalt} }%
	{%
	\put(\xCoord,#2){\line(1,0){\masslinelen} \raisebox{#4}{\hspace{#3}$#1$}}%
	}%
	{%
	\put(\xCoord,#2){\line(1,0){\masslinelen} \raisebox{#4}{\hspace{#3}$#1$}}%
	\gobblemasslist#5\gobbleEnd%
	}%
}%

\newlength{\oldunitlength}
\setlength{\oldunitlength}{\the\unitlength}
\newcommand{\masslinelen}{50}
\newcommand{\xCoordColOne}{10}
\newcommand{\xCoordColTwo}{120}
\newcommand{\xCoordColThree}{230}
\newcommand{\xCoordColFour}{340}
\newcommand{\xCoordColFive}{450}
\newcommand{\xCoord}{}
\newcommand{\defaultX}{1mm}
\newcommand{\defaultY}{-0.6ex}
\newcommand{\placetoleft}{-1.4cm}
\newcommand{\placetoleftnudgeright}{-1.33cm}
\newcommand{\slightlyraised}{-0.4ex}
\newcommand{\slightlylowered}{-0.8ex}


\newcommand{\E}[1]{\times 10^{#1}}	
\newcommand{\vev}[1]{\inap{#1}}		

\newcommand{\MP}{M_{\text{Pl}}} 		
\newcommand{\MSUSY}{M_\text{SUSY}}	
\newcommand{\Mmess}{M_\text{mess}}	

\newcommand{\G}[1]{\ensuremath{G_{#1}}}	
\newcommand{\SUSYLR}{SUSYLR}
\newcommand{\CISLR}{CISLR}

\newcommand{\WSUSYLR}{W_{\text{\SUSYLR}}}

\newcommand{\ourmodel}{AMSBLR}

\newcommand{\DeltaBar}{\bar{\Delta}}

\newcommand{\DeltaC}{\Delta^c}
\newcommand{\DeltaBarC}{\bar{\Delta}^c}

\newcommand{\scalar}[1]{%
\ifthenelse{ \equal{#1}{\DeltaC} }%
	{%
	\underline{\Delta}^c%
	}%
{ \ifthenelse{ \equal{#1}{\DeltaBarC} }%
	{%
	\underline{\bar{\Delta}}^c%
	}%
{ \ifthenelse{ \equal{#1}{\Phi} }%
	{%
	\underline{\Phi}%
	}%
	{%
	\underline{#1}%
	}}}%
}%

\newcommand{\DBarCpp}{\bar{\Delta}^{c ++}}
\newcommand{\DBarCp}{\bar{\Delta}^{c +}}
\newcommand{\DBarCz}{\bar{\Delta}^{c \,0}}
\newcommand{\DCmm}{\Delta^{c --}}
\newcommand{\DCm}{\Delta^{c -}}
\newcommand{\DCz}{\Delta^{c \,0}}
\newcommand{\phiUp}{\Phi_u^+}
\newcommand{\phiUz}{\Phi_u^0}
\newcommand{\phiDz}{\Phi_d^0}
\newcommand{\phiDm}{\Phi_d^-}

\newcommand{\Slight}{N}
\newcommand{\Sheavy}{S}

\newcommand{\kappaLight}{\kappa_\Slight}
\newcommand{\kappaHeavy}{\kappa_\Sheavy}

\newcommand{\lambdaLight}{\lambda_\Slight}
\newcommand{\lambdaHeavy}{\lambda_\Sheavy}

\newcommand{\I}{\mathbbm{i}}
\newcommand{\EE}{\mathbbm{e}}

\newcommand{\bibpath}{.}
\bibliographystyle{h-physrev}
\citestyle{plain}

\begin{document}

\maketitle

\section{Acknowledgements}
\fi%

We are indebted to Zackaria Chacko for his discussion and proof-reading of our paper.  We would also like to thank Markus Luty and Ken Hsieh for helpful discussion on AMSB.  Furthermore, Michael Ratz has our appreciation for his discussion on the early universe vacua.  Finally we would like to acknowledge the assistance of Craig Group for help with the online tool SUPERSIM.  This work was supported by the National Science Foundation grant no.~Phy-0354401.  

\ifglobal{}%
\else%
%


\end{document}
\fi%

%

\bibliography{%
\bibpath/amsb,%
\bibpath/susy,%
\bibpath/susylr,%
\bibpath/susy-extended_higgs,%
\bibpath/nmssm,%
\bibpath/susy-ext_to_superspace,%
\bibpath/flavor_violation,%
\bibpath/computer_codes,%
Pheno,%
CISLR,%
Intro%
}


\end{document}